\documentclass[prc,twocolumn,floatfix,groupedaddress,showpacs,superscriptaddress] {revtex4}
\usepackage{graphicx}% Include figure files
\usepackage{dcolumn}%Align table columns on decimal point
\usepackage{mathrsfs}
\usepackage{bm}% bold math
\usepackage{dsfont}
\usepackage{dcolumn}% Align table columns on decimal point
\usepackage[usenames]{color}
\usepackage{amssymb}
\usepackage{amsmath} % Align equations
\usepackage{makeidx}
\usepackage{array}
\usepackage{multirow}
\usepackage{bm} %bold math
\usepackage{soul} % crossing out text
\usepackage[normalem]{ulem}
\DeclareMathAlphabet{\mathpzc}{OT1}{pzc}{m}{it} %define calligraphic font for small
\usepackage{color}

\definecolor{OliveGreen}{rgb}{0,0.6,0}

\begin{document}
\title{Ab initio study of resonance states of exotic $^{10}$Li nucleus}

\author{D. M. Rodkin}
\affiliation{Dukhov Research Institute for Automatics, 127055, Moscow, Russia}

\affiliation{Moscow Institute of Physics and Technology, 141701 Dolgoprudny, Moscow Region, Russia
}
\author{Yu. M. Tchuvil'sky}
\affiliation{Skobeltsyn Institute of Nuclear Physics, Lomonosov Moscow State University,
119991 Moscow, Russia}
\date{\today}

\begin{abstract}
The spectrum, partial and total decay widths of the states of the exotic $^{10}$Li nucleus are studied  in an ab initio approach. The spectrum is calculated using No-Core Shell Model and corresponding extrapolation procedure. The calculations use the well-proven Daejeon16 potential. The NCSM-based approach, which includes a method for constructing the basis of functions of cluster channels and a procedure for matching the cluster form factors obtained within this method with the asymptotic wave functions, is applied to compute the widths of the  levels. The results obtained quite adequately reproduce the pattern given by the most advanced experiments. Detailed predictions of the spectral characteristics of $^{10}$Li nucleus states are given.
\end{abstract}

\pacs{}
\maketitle

The study of the properties of atomic nuclei lying in the drip-line region (both bound and unbound)  is one of the most interesting and popular problems of modern low-energy nuclear physics. Same as $^5$He and $^7$He nuclei $^{10}$Li nucleus  is a very interesting unbound system located between two isotopes having bound ground states. Additional interest in this system is attracted by the fact that contrary to the He isotopes $^{10}$Li is odd-odd nucleus. Equally intriguing is that $^{10}$Li is a subsystem of $^{11}$Li nucleus that is unique in its properties. Thus the study of $^{10}$Li may answer the question of the presence or absence of inversion in $^{11}$Li nucleus of $2s_{1/2}$ and $1p_{1/2}$ orbitals. 

To date, a lot of experiments have been carried out, the purpose of which was to study $^{10}$Li resonances. Many reactions were used. Early, up to 2004, experimental results can be found in paper \cite{exp10li}. The pattern of experimental data presented in this review looks quite mixed and intricated. It has been fairly reliably established that: a) the n+$^9$Li $s$-wave makes a noticeable contribution to the cross sections for most reactions at low energies, b) there is a pronounced resonance (most likely in the $p$-wave, i.e., this resonance corresponds to the 1$^+$ or 2$^+$ state) in the energy range 0.2 -- 0.6 MeV. It is also useful to note  that all the data on the decay widths collected in the review point to the relative smallness of these widths ($\Gamma < $  0.5 MeV).  The sole exception is the result of the first observation of the $^{10}$Li system, in which it was identified using reaction $^9$Be($^9$Be,$^8$B)$^{10}$Li \cite{first}. The peak with the width $\Gamma$=1.2$\pm$ 0.3 MeV was observed at the energy E=0.80$\pm$ 0.25 MeV.

In subsequent years, studies of $^{10}$Li in the processes of stopped pion absorption (see \cite{e5} and Refs. therein)
were carried out. One- and two-proton removal  reactions from $^{11}$Be \cite{e52} and  $^{12}$B \cite{e52,e6} were also used. Of greatest interest among the experiments under discussion are, in our opinion, those in which processes relatively simple in their mechanism are used, namely, the reactions of removal and pick-up of one neutron p($^{11}$Li,d(or pn))$^{10}$Li \cite{e7,e8,e9,e10} and d($^9$Li,p)$^{10}$Li  \cite{e11,e12,e13,e14}, respectively. These experiments provided a variety of information about the resonance states of $^{10}$Li nucleus. Almost all of the experiments performed confirmed the previous data on the presence of $p$-wave resonance in the region of 0.5 MeV and a noticeable contribution of the $s$-wave to the continuous $^9$Li + n spectrum at low energies. In many papers, the smallness of the width of the $p$-wave resonance was also confirmed. The question of the absence or presence, as well as the positions and widths of higher-lying resonances, these works have not fully clarified. From some of their results and their processing in the theoretical papers cited below, only the strong influence of the $d$-wave on the properties of the continuum in the region E $>$1.5 MeV was reliably established. 

Among the works discussed, Ref. \cite{e14} stands apart, in which the results of studying the states of $^{10}$Li in reaction d($^9$Li,p)$^{10}$Li at the beam energy of 100 MeV are presented. The main advantage of this experiment is the  great beam intensity, which made it possible to achieve almost 2 orders of magnitude higher statistics compared to the other experiments. The $^{10}$Li energy spectrum is measured up to 4.6 MeV. As a result of the analysis of the obtained data, it was concluded by the authors that there exist three resonances with parameters: $ E_1 =0.45 \pm 0.03$ MeV $\Gamma_1=0.68 \pm 0.03$ MeV, $ E_2 =1.5 \pm 0.1$ MeV $\Gamma_2=1.1 \pm 0.3$ MeV, and $ E_3 =2.9 \pm 0.3$ MeV $\Gamma_3=2.6 \pm 0.6$ MeV in the spectrum of $^{10}$Li. The first of them is generated by the $p$-wave, the second is a mixture of $p$- and $s$-waves, and in the third, the $d$-wave is dominant.

Naturally, the primary task of theoretical studies of unstable states of  $^{10}$Li was to describe the reactions in which these states are populated. The results of these studies are contained both in works describing experiments, for example, in Ref. \cite{e6} and in separate theoretical works devoted to the discussed subject.  The ordinary shell-model estimates \cite{t5,t6}, the coupled-channel method \cite{e13,t01,t02}, its microscopic version \cite{t03}, time-dependent projectile fragmentation model \cite{t1}, DWBA \cite{e10,t2},  transfer to the continuum formalism \cite{t3,t4}, theory of quasi-elastic knock-out reactions \cite{e14}, induced surrogate
reaction formalism \cite{t8}, etc. were used. Almost all these works operate with the concepts of partial waves of relative motion $^{10}$Li + n, and not with the characteristics of the channel spin and the total angular momentum of the ten-nucleon system. They provide reliable confirmation of the above experimental results: the presence of a low-lying resonance and the contribution of both $p_{1/2}$- and $s_{1/2}$-waves in this energy range. A special place among these papers is occupied by Ref. \cite{t3}, which analyses the most statistically significant results of Ref. \cite{e14}, and also demonstrates the absence of contradictions of these results with the data presented in Ref. \cite{e13}. The difference from the Lorentzian shape of the peak in the cross section found in the latter work is explained by the contribution of the $s$-wave. As a result, this paper confirms the data on the position of this resonance (E $\sim$ 0.45 MeV) from Ref. \cite{e14}. The presence of a resonance in the energy range E $>$ 2 MeV where the $d$-wave dominates is also confirmed. Another important confirmation of the previous experimental data was presented in Ref. \cite{t8}, which is discussed bellow in more details.

The studies of the excitation functions and differential cross sections of the reactions under discussion is based on calculations of the structure of the ten-nucleon system and the nuclei entering into reactions to produce it. For these purposes, three- and two-particle calculations of $^{11}$Li and $^{10}$Li nuclei within the potential model  (including Faddeev ones)  \cite{t01}, standard shell model \cite{t5,t6}, renormalized nuclear field theory  \cite{t8}, Hartree-Fock-Bogolyubov-Gorkov approach \cite{t2}, etc. were used. The coupled-channel method was also applied to describe the structure of the reacting nuclei  \cite{t03}.

To calculate the spectrum of the $^{10}$Li nucleus, advanced microscopic approaches, such as the Generator Coordinate  Method (GCM) \cite{ts9} and  Microscopic Multicluster Model (MMM) \cite{ts10} were exploited. In the former work, the (reduced) decay widths of $^{10}$Li states were computed too. Ab initio approaches were also used to study the spectrum, namely No-Core Shell Model (NCSM) \cite{ts11} and Gamov Shell Model (GSM) \cite{ts12}. However, only the level energies were computed in these works. 

In the light of the above, the task of simultaneously describing the energies and decay widths of $^{10}$Li nucleus states within the framework of an ab initio approach seems to be timely and acutely relevant. In this paper we demonstrate a solution to this problem using developed by us the Cluster Channel Orthogonalized Functions Method (CCOFM) \cite{our1,our2,our3,our4}. The last two papers present the results of calculations of the total binding energies (TBEs) of the $^8$Be and $^7$Li nuclei, their spectra in a wide range of excitation energies, the partial decay widths of the states of these nuclei into nucleon and cluster channels, etc. These papers provide a detailed description of the method used, justify its reliability and demonstrate the successful description of known experimental data. Paper \cite{our5}  is devoted to the study of $^7$He nucleus, all states of which are unstable. The spectroscopic characteristics of all such states up to 5 MeV are predicted. In terms of the level of detail of the obtained data, the results of the discussed work significantly exceed similar results of both experimental and theoretical works devoted to $^7$He nucleus. 

In the present paper we demonstrate the results of application of CCOFM to the detail study of the states of $^{10}$Li nucleus. The approach includes: M-scheme NCSM calculations of the parent and daughter lithium isotopes, extrapolation of the obtained binding energies of these nuclei, projecting of the wave functions of $^{10}$Li states  into various $^9$Li + n channels, and the procedure of matching the resulting projections -- cluster form factors (CFF) -- with the corresponding asymptotic channel functions.

In the NCSM computations the Daejeon16 potential \cite{dj16} is exploited as a model of NN-interaction. 
It is built using the N3LO limitation of Chiral Effective Field Theory \cite{ceft1} softened by Similarity
Renormalization Group transformation \cite{srg1}. This potential is designed to calculate 
all kinds of characteristics of nuclei with the masses $A \leq 16.$ It has been tested in many studies of various properties of nuclei. 
These tests demonstrates that such characteristics are, in general, reproduced well.  This choice is supported by our previous studies of the decay properties of light nuclei \cite{our3, our4,our5}. The NCSM calculations were carried out with the use of the Bigstick code \cite{bigstick}.  In the computations of the TBEs of $^{10}$Li the bases are limited by the values of cut-off parameter $N^{*max}_{tot}$=10 and 9 for the positive and negative parity states respectively. TBEs of two lower states of the fragment $^9$Li  are also computed using the basis with  $N^{*max}_{tot}$=10.  

Five-parameter "Extrapolation A5"  method \cite{extrapA} is used for obtaining TBEs of $^{10}$Li and $^9$Li states in "infinite" shell-model basis. The values of the parameters are determined for each level independently by fitting the extrapolating function  to the theoretically calculated TBEs.
We fit these parameters  in the $\hbar \omega$ range from 10 to 25 MeV with 2.5 MeV step and all available dimensions of the bases. It turned out that $\hbar \omega$ = 15 MeV is the optimal value for lower levels of $^{10}$Li and $^9$Li, so the wave functions calculated using this value are chosen for the computations of the widths.

It should be pointed out that in this work, we give preference and use only the modern (so-called "new") definition of the CFF \cite{newsf1,newsf2,newsf3} (see just mentioned papers and the discussion in Ref. \cite{our4}). Therefore, only the data obtained in this scheme are presented below.

The CFF $\Phi^{c_\kappa}_A(\rho)$  describes the relative motion of subsystems in A-nucleon configuration space. In the modern definition, CFF is the following overlap
$$\Phi^{c_\kappa}_A(\rho)=$$
\begin{equation}
\langle \Psi _{A}|\hat A\{\Psi^{\{k_1\}} _{A\,_1} \Psi^{\{k_2\}}
_{A\,_2}\hat N^{-1/2}\frac{\delta(\rho-\rho')}{\rho'^2}Y_{l} (\Omega)\} _{J_c,M_JT}\rangle ,\label{cff}
\end{equation} 
where $\Psi _{A}$  is the wave function (WF) of the initial nucleus -- the  NCSM solution of the A-nucleon Schr\"odinger equation,  $\hat A$ is the antisymmetrizer, $\Psi^{\{k_i\}}_{A\,_i}$ is a translationally invariant internal WF of the fragment labelled by a set
of quantum numbers $\{k_i\}$; $\varphi _{nlm} (\vec \rho )$ is the WF of the relative
motion. The channel WF as a whole is labelled by the set of quantum numbers $c_\kappa$ that
includes $\{k_1\},\{k_2\},l,J_c,M_{J_c},M_J,T$, where $J_c$ is
the channel spin. $\hat N$ is the norm operator of the generalized function of the cluster channel which takes the form:

$$\hat N(\rho', \rho'')=\langle \hat A\{\Psi^{\{k_1\}} _{A\,_1} \Psi^{\{k_2\}}
_{A\,_2}\frac{\delta(\rho-\rho')}{\rho'^2}Y_{l} (\Omega)\} _{J_c,M_JT}|\times$$
\begin{equation}
|\hat A\{\Psi^{\{k_1\}} _{A\,_1} \Psi^{\{k_2\}}
_{A\,_2}\frac{\delta(\rho-\rho'')}{\rho''^2}Y_{l} (\Omega)\} _{J_c,M_JT}\rangle.\label{norm}
\end{equation} 
The spectroscopic factor (SF) is the CFF norm.

Representation of the generalized function of the relative motion in the form of an expansion in terms of oscillator functions
\begin{equation}
[\delta(\rho-\rho')/\rho'^2]Y_{lm} (\Omega)=\sum\limits_n\varphi _{nlm} (\vec \rho )\varphi _{nlm} (\vec \rho' )\label{delta}
\end{equation}
reduces the norm operator to the overlap kernel matrix and makes it possible to write the CFF in the form 
 
\begin{equation}
\Phi^{c_\kappa}_A(\rho)=\sum\limits_k \varepsilon^{-1/2} _{\kappa,k} \langle \Psi _{A}|\hat A\{ \Psi^{\{k_1\}} _{A_1}
\,\Psi^{\{k_2\}}_{A_2} \,f_l^k (\rho') \} \rangle f_l^k (\rho ). \label{eqcffn}
\end{equation}
The eigenvalues and eigenvectors of this overlap kernel matrix can be written as:

\begin{equation}
\varepsilon _{\kappa,k}  =  \langle \hat A\{ \Psi^{\{k_1\}} _{A_1} \,\Psi^{\{k_2\}} _{A_2} \,f_l^k (\rho
) \} |\hat 1|\hat A\{ \Psi^{\{k_1\}} _{A_1} \,\Psi^{\{k_2\}}_{A_2} \,f_l^k (\rho
) \}  \rangle ; \label{eq8}
\end{equation}

\begin{equation}
f_l^k (\rho ) = \sum\limits_n {B_{nl}^k \varphi _{nl} (\rho )}. \label{eq9}
\end{equation}
 
 After that the CCF can be expressed in the form of an expansion in the oscillator basis:
\begin{equation}
\Phi^{c_\kappa}_A(\rho)=\sum\limits_k\varepsilon^{-1/2} _{\kappa,k}\sum\limits_{n, n'}B_{nl}^k B_{n'l}^{k}
C_{AA_1A_2}^{n'l}\varphi _{nl} (\rho ) \label{eq11}
\end{equation}
 The coefficient contained in this expression has the form
 \begin{equation}
C_{AA_1A_2}^{nl} = \langle \hat A\{\Psi^{\{k_1\}} _{A_1} \Psi^{\{k_2\}} _{A_2} \varphi _{nl}
(\rho ) \} | \Psi _{A} \rangle
\end{equation}
and called the spectroscopic amplitude. Methods of its ab initio calculation are described in Refs. \cite{our1, our2}. 

\begin{figure}[htp]
\includegraphics[width=0.5\textwidth]{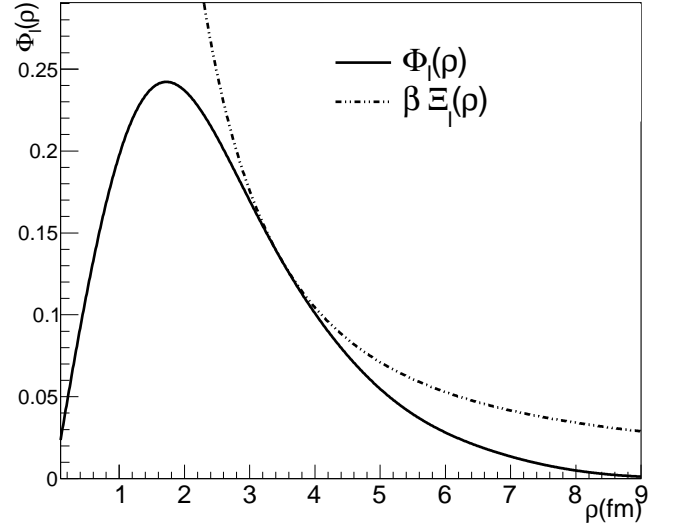}
\caption{Picture of matching of CFF of $1^+$ state and the asymptotic function in the channel with $l(S)= 1(2)$.}\label{fig1}
\end{figure}

As in our previous works \cite{our4,our5}, we use the procedure of matching the CFF with the asymptotic wave function of the corresponding channel. In general case (for both charged and neutral emitted particles)  the procedure includes, first, finding points $ \rho_{m}$ in which the logarithmic derivatives of  CFF and function

\begin{equation}
\Xi_l ( \rho)  = (F^2_l ( \rho) + G^2_l ( \rho))^{1/2} \label{log}
\end{equation}
are coincide. Second, by comparing the values of the functions at one of these points, the amplitude of the decay channel function at infinity is determined. The width of a resonance is calculated using the conventional R-matrix theory:
\begin{equation}
\Gamma  = \frac{\hbar^2}{\mu k_0} \Xi_l ( \rho)^{-2}(\Phi _A^{c_\kappa} (\rho_{m} ))^2. \label{gam}
\end{equation}
A typical result of such a procedure is illustrated by Fig.\ref{fig1}. 
 So the CFF in its new definition allows matching with the asymptotic wave function at relatively small distances. Close coincidence of the matched functions takes place in a fairly wide ($\sim$1 fm) domain. The deviation of these functions from each other in this domain turns out to be less than 5 $\%$. Such a deviation, obviously, cannot seriously affect the conclusions of this work.

So, in this paper the approach of Refs. \cite{our1,our2,our3,our4,our5} is used to study the spectrum of states of the $^{10}$Li nucleus and the decay widths of these states into various channels.

The TBEs, excitation energies $E^*$, as well as decay energies $E^n$, SFs, and widths of the levels of $^{10}$Li system are calculated. The energy of the first resonant state $1^+$ turns out to be equal to $E^n$=0.95 MeV. Taking into account that it is hardly possible to achieve an accuracy of the computed result higher than 1 MeV for the energy of resonant state of an exotic nucleus, we, basing on the result of experimental work \cite{e14} where the value $E^n$=0.45 MeV was obtained, have used this value in calculating its widths. This technique is close in meaning to the NCSMC-pheno ansatz presented in  \cite{fen1}. Similar corrections were used in our previous works \cite{our4, our5}. What about all other obtained levels, experimental data on their positions are contradictory and cannot be used to calculate decay widths. Therefore, the sole phenomenological correction of the level energies calculated by us, namely their shift $\Delta E=-0.50$ MeV, analogous to the shift of the lowest one has been introduced.

The results of NCSM computations obtained within the framework of the just described procedure are presented in Tab. I. In  the energy range of about 4.5 MeV above the neutron threshold 12 resonant states are found including several second levels with given $J^{\pi}$ values. Such a large list of levels is obtained for the first time. In previous NCSM calculations of the spectrum of $^{10}$Li \cite{ts11} the effective interaction derived from the SD-Bonn 2000 NN-potential was used. Almost degenerate states of positive parity $1^+$, $2^+$, and $0^+$ were found to be localized just above the threshold. The levels $2^-$, $1^-$, and $0^-$ were also found at energies of about 3, 4, and 5 MeV. In the framework of GSM \cite{ts12}, the states $2^+$, $1^+$, $1^-$, and $2^-$ with energies of 0.35, 0.68, 1.05, and 1.05 MeV, respectively, were obtained. A spectrum obtained in MMM \cite{ts10} includes $1^+$, $2^+$, $2^-$, $1^-$, and $0^+$ states at energies 0.39, 0.55, 1.56, 1.74 and 2.12 MeV.  The levels $1^+$, $2^+$, and $0^+$ computed in the framework of the GCM with the Volkov2 forse in Ref. \cite{ts9} are placed at energies  $- 0.30$, 0.86, and 1.9 MeV. The widths of unbound ones are not large.  According to the author, the existence  or absence of a broad virtual $s$-state  is rather model dependent.

\begin{center}
\begin{table}
\caption{Resonance energies (MeV), SFs, decay widths of channels characterized by angular momentum $l$ and channel spin $S$ (keV) of $^{10}$Li nucleus states. The subscripts denote the mass of Li isotopes.}
\begin{tabular*}{0.4\textwidth}{ c c c c c c c c c }
\hline\hline\noalign{\smallskip}
$J^{\pi}_{10}$  &$E^*_{10}$  &$J^{\pi}_9$ &$E^*_9$&$E^n$ & $l(S)$ &SF &$\Gamma^n$ \\

\hline\noalign{\smallskip}  

1$^{+}_1$  & 0 & 3/2$^-$ &0& 0.45 & \begin{tabular}{c} 1(1)\\1(2) \end{tabular}&\begin{tabular}{c} 0.805\\ 0.800\end{tabular} &\begin{tabular}{c} 281\\ 266\end{tabular} \\ 

\hline\noalign{\smallskip}  

2$^{-}_1$& 0.44 & 3/2$^-$ &0 & 0.89 & \begin{tabular}{c} 2(1)\\2(2)\\0(2) \end{tabular}&\begin{tabular}{c} 0.072\\ 0.071\\0.752\end{tabular} &\begin{tabular}{c} 6.7\\ 7.0\\1670\\ \end{tabular}& \\ 

\hline\noalign{\smallskip}  

2$^{+}_1$ & 0.47 & 3/2$^-$&0 & 0.92 & \begin{tabular}{c} 1(1)\\1(2)\\3(2) \end{tabular}&\begin{tabular}{c} 0.688\\ 0.845\\0.0026\end{tabular} &\begin{tabular}{c}542\\ 681 \\14.7 eV\end{tabular}&\\ 

\hline\noalign{\smallskip}  

1$^{-}_1$  & 1.02 & 3/2$^-$&0 & 1.47 & \begin{tabular}{c} 2(2)\\0(1) \end{tabular}&\begin{tabular}{c} 0.039\\ 0.804\end{tabular} &\begin{tabular}{c}  19\\ 2560\end{tabular}&\\ 

\hline\noalign{\smallskip}  

0$^{+}_1$ &1.69  &  3/2$^-$  & 0  & 2.14  & 1(1) & 0.650 & 1388&\\ 

\hline\noalign{\smallskip}  

1$^{-}_2$  & 2.34  &\begin{tabular}{c}  3/2$^-$\\ \\ \\ 1/2$^-$ \\ \end{tabular} &\begin{tabular}{c} 0\\ \\ \\2.69 \\ \end{tabular}  &\begin{tabular}{c}2.79 \\ \\ \\0.10\\ \end{tabular}&  \begin{tabular}{c} 2(1)\\2(2) \\0(1) \\0(1) \\ \end{tabular}&\begin{tabular}{c} 0.395\\0.202 \\0.056 \\0.394 \\ \end{tabular}  &\begin{tabular}{c} 392\\246\\223\\148 \\ \end{tabular}&\\ 

\hline\noalign{\smallskip}  

2$^{+}_2$ &3.07&\begin{tabular}{c}  3/2$^-$\\ \\1/2$^-$\end{tabular}  &\begin{tabular}{c} 0\\ \\2.69\end{tabular} & \begin{tabular}{c} 3.52\\ \\0.83 \end{tabular} & \begin{tabular}{c} 1(1)\\1(2) \\1(1) \end{tabular}&\begin{tabular}{c} 0.088\\ 0.004 \\0.008\end{tabular} &\begin{tabular}{c} 242\\17\\ 5.5\end{tabular}&\\ 

\hline\noalign{\smallskip}  

1$^{+}_2$  & 3.25 &\begin{tabular}{c}   3/2$^-$\\ \\ \\1/2$^-$ \\ \\\end{tabular}  &\begin{tabular}{c} 0\\ \\ \\2.69 \\ \\\end{tabular}  &\begin{tabular}{c}3.70 \\ \\ \\1.01 \\ \\ \end{tabular}& \begin{tabular}{c} 1(1)\\1(2) \\ 3(2)\\1(0) \\1(1) \end{tabular}&\begin{tabular}{c} 0.032\\ 0.005 \\0.008 \\0.403 \\0.735\end{tabular} &\begin{tabular}{c} 122\\23.2\\3.8\\377\\ 658\\\end{tabular}&\\ 

\hline\noalign{\smallskip}  

4$^{-}_1$& 3.34 &\begin{tabular}{c}   3/2$^-$\\ \\ \\ 1/2$^-$\\ \\ \end{tabular} &\begin{tabular}{c} 0\\ \\ \\ 2.69 \\ \\ \end{tabular}  &\begin{tabular}{c}3.79 \\ \\ \\1.10\\ \\ \end{tabular}& \begin{tabular}{c} 2(2)\\ 4(1)\\4(2)\\ 4(0)\\4(1)  \end{tabular}&\begin{tabular}{c} 0.852\\0.002 \\0.001\\ 0.002\\0.004 \end{tabular} &\begin{tabular}{c} 1410 \\ 0.13\\0.023\\ 0.23 eV\\1.2 eV\end{tabular}&\\ 

\hline\noalign{\smallskip}  

0$^{-}_1$ & 3.35  &\begin{tabular}{c}   3/2$^-$\\ 1/2$^-$\end{tabular}  &\begin{tabular}{c} 0\\ 2.69\end{tabular}  &\begin{tabular}{c}3.80\\1.11 \end{tabular}& \begin{tabular}{c} 2(2)\\ 0(0) \end{tabular}&\begin{tabular}{c} 0.163\\ 0.739\end{tabular}&\begin{tabular}{c} 336\\1940\end{tabular}&\\ 

\hline\noalign{\smallskip}  

3$^{+}_1$ & 3.64 &\begin{tabular}{c}   3/2$^-$\\ \\ \\1/2$^-$\end{tabular}&\begin{tabular}{c} 0\\ \\ \\2.69 \end{tabular}  &\begin{tabular}{c} 4.09 \\ \\ \\1.40\end{tabular}& \begin{tabular}{c} 1(2)\\3(1) \\ 3(2)\\3(0) \end{tabular}&\begin{tabular}{c}0.495 \\ 0.001 \\0.001 \\0.002\end{tabular} & \begin{tabular}{c} 1754\\0.49\\0.53\\37 eV\end{tabular}&\\ 

\hline\noalign{\smallskip}  

2$^{-}_2$  & 4.00 &\begin{tabular}{c}  3/2$^-$\\ \\ \\ \\1/2$^-$\\ \\ \end{tabular}  &\begin{tabular}{c} 0\\ \\ \\ \\2.69\\ \\ \end{tabular} &\begin{tabular}{c}4.45 \\ \\ \\ \\1.76\\ \\  \end{tabular}& \begin{tabular}{c} 2(1)\\2(2) \\4(2)\\0(2) \\2(0)\\2(1)  \end{tabular}&\begin{tabular}{c} 0.330\\ 0.248 \\0.002 \\ 0.254\\0.025\\0.064 \end{tabular} &\begin{tabular}{c} 706\\525\\0.17\\1519\\8.8\\28.2\end{tabular}&\\

\hline\noalign{\smallskip}
\end{tabular*}

\end{table}
\end{center}

An essential feature of the spectrum  presented in Tab. I, is that all the levels other than 1$^{+}_1$ and 2$^{+}_2$ have a great total decay width. In this respect, the results differ sharply from the results of most experimental and theoretical studies. The exceptions are the data of  \cite{e14} (see above) and the results of latest theoretical works  \cite{t3,t4}.

Analysis of the properties of the obtained spectrum leads to the following conclusions. The existence of a pronounced 1$^{+}$  $p$-wave resonance is confirmed. The calculated value of the total decay width of it which is equal to  547  keV is in a good agreement with the experimental one presented in Ref. \cite{e14}, wherein it should be taken into account that the contribution to the cross sections of the reactions shaping this state also comes from a neighboring broad resonance  $2^+_1$. Large partial decay widths of negative-parity states $2^-_1$, $1^-_1$, and $2^-_2$ into decay channels $^9$Li+n with $l$=0 demonstrate that the $s$-wave makes a significant contribution to the yield probability of these products throughout the studied energy area. Likewise, the presence of wide resonance states with $l$=1 shows that the $p$-wave has similar properties. The contribution of the $d$-wave to this probability is determined by the population or decay of the states $1^-_2$, $4^-_1$, $0^-_1$, and $2^-_2$. It becomes noticeable starting from an energy of $\sim$ 2.5 MeV. This effect is also well confirmed by the data of various experiments, in particular \cite{e14}, as well as the related theoretical studies.

A detailed comparison of the spectrum obtained by us with the data of Ref. \cite{e14} demonstrates their very good agreement. Indeed, in place of the second 1.5 MeV ``resonance'' identified in the discussed work, in which $s$- and $p$-waves are almost equally represented (here we reproduce the terminology of Ref. \cite{e14}, somewhat different from that used by us), our calculations indicate the presence of a wide $s$-resonance $1^-_1$, overlapping with neighboring bellow and above $p$-resonances $2^+_1$ and $0^+_1$ (according to  Ref. \cite{e14} the cross section maximum is observed at $E_{max}$ = 1.9 MeV). The third identified resonance at 2.9 MeV (the corresponding maximum is placed at $E_{max}$ = 3.9 MeV) which is dominated by the $d$-wave is, most likely, an overlap of resonances  $1^-_2$, $4^-_1$, $0^-_1$, and $2^-_2$. 

The results of the theoretical analysis of the data of Ref. \cite{e14} presented in Ref. \cite{t8} are also correlate well with ours. The maxima and widths of the strength functions of $^{10}$Li states, deduced in this paper for the d($^9$Li,p)$^{10}$Li reaction at 100 MeV
incident energy, are nearly the following: $1^+$ -- 0.35 MeV and 0.5 MeV, $2^+$ -- 0.75 MeV and 1.0 MeV, $4^-$ -- 3.7 MeV and $\geq$1.5 MeV.  Weakly expressed broad maxima of strength functions of resonances $3^-$, $2^-$, and $1^-$, as well as of $p_{3/2}$-wave were also found. The only difference is the absence of state  $3^-$ in the spectrum obtained by us.

The data presented in other experimental works contradict each other to one degree or another. As stated above, earlier works  usually show resonances of small width and more often identify the ground state with a negative parity virtual state (or resonance), or with a doublet of this state and $1^+$ one. In later papers, including those mentioned above, resonances are sometimes indicated for which $\Gamma \sim E$. The ground state of $^{10}$Li is usually interpreted as a $p$-resonance, i. e. no parity inversion occurs. Our results indicate, rather, the absence of inversion.  At the same time, difference $E_{1^+}-E_{2^-}=0.44$ MeV for the extrapolated energy values  is not too large, so it cannot be ruled out that when using another realistic model of the internucleon interaction, the inversion will be obtained. So our results do not allow one to draw an unambiguous conclusion concerning the parity inversion.

The results presented in this work predict various effects that can manifest themselves in the processes of population and decay of $^{10}$Li states. The most interesting, in our opinion, are the processes of resonant excitation of the  1/2$^{-}$  $E^*$=2.69 MeV state of $^{9}$Li nucleus. One can expect large cross sections of these processes as a consequence of the great partial widths of the corresponding $^{10}$Li population-decay channels. For example, reaction  d($^9$Li,pn)$^9$Li looks promising to observe that. 

As a conclusion, we point out that in this paper we report results of the first ab initio computation of the complete set of $^{10}$Li nucleus spectral characteristics  up to 4.5 MeV. The calculated spectrum turns out to be a set of overlapping broad resonances of different spin and parity. The results correlate well with the most statistically significant modern experimental data and thus create a complete and consistent description of the spectral properties of $^{10}$Li nucleus. The findings of our investigation can serve as a basis for predicting the results of measuring the cross sections for various processes of population and decay of this actively debated and studied object. 

The study is supported by a grant from the Russian Science Foundation No. 22-22-00096, https://rscf.ru/en/project/22-22-00096/.
We are grateful to R. Wolski for valuable information, to A. M. Shirokov for providing the Daejeon16 NN-potential matrixes, as well as  to C. W. Johnson for supporting our efforts to introduce code Bigstick for NCSM calculations.

\end{document}